\documentclass[namedreferences]{solarphysics}

\usepackage[hyperref,optionalrh,showbiblabels]{spr-sola-addons} % For Solar Physics 
\usepackage{graphicx}        % For eps figures, newer & more powerfull
\usepackage{color}           % For color text: \color command
\usepackage{breakurl}        % For breaking URLs easily trough lines
\usepackage{natbib}
\usepackage{enumerate}
\usepackage{float}
\usepackage[utf8]{inputenc}
\chardef\us=`\_

\begin{document}

\begin{article}
\begin{opening}

\title{A Multiwavelength Analysis of the Long-duration Flare Observed on 15 April 2002}

\author[addressref=aff1,corref,email={ak@cbk.pan.wroc.pl}]{\inits{A.}\fnm{A.}~\lnm{Kepa}\orcid{}}\sep
\author[addressref=aff1,email={bs@cbk.pan.wroc.pl}]{\inits{B.}\fnm{B.}~\lnm{Sylwester}\orcid{}}\sep
\author[addressref=aff1,email={js@cbk.pan.wroc.pl}]{\inits{J.}\fnm{J.}~\lnm{Sylwester}\orcid{}}\sep
\author[addressref={aff1,aff2},email={tmrozek@cbk.pan.wroc.pl}]{\inits{T.}\fnm{T.}~\lnm{Mrozek}\orcid{}}\sep
\author[addressref=aff1,email={ms@cbk.pan.wroc.pl}]{\inits{M.}\fnm{M.}~\lnm{Siarkowski}\orcid{}}

%{Space Research Center, Polish Academy of Sciences, Kopernika 11, 51-622 Wroclaw, Poland}
\address[id=aff1]{Space Research Centre (CBK PAN), Bartycka 18A, Warsaw,  Poland}
\address[id=aff2]{Astronomical Institute, University of Wroclaw, ul. Kopernika 11, Wroclaw, Poland}

\runningauthor{A. Kepa \textit{et al.}}
\runningtitle{A Multiwavelength Analysis of the Long-duration Flare Observed on 15 April 2002}

\begin{abstract}
We present a multiwavelength analysis of the long duration flare observed on 15 April 2002 (soft X-ray peak time at 03:55 UT, {\textsf{SOL2002-04-15T03:55}}). This flare occurred on the disk (S15W01) in NOAA 9906 and was observed by a number of space instruments including the \textit{Extreme-Ultraviolet Imaging Telescope} on the \textit{Solar and Heliospheric Observatory} (SOHO/EIT), the RESIK spectrometer onboard the \textit{Coronas-F} spacecraft, and the \textit{Ramaty High Energy Solar Spectroscopic Imager} (RHESSI). We have performed a complex analysis of these measurements and studied the morphology and physical parameters characterizing the conditions in flaring plasmas. The~195~\AA~SOHO/EIT images have been used to study evolution of flaring loops. Analysis of RHESSI data provided the opportunity for a detailed analysis of hard X-ray emission with~1~keV energy resolution. We have used \textit{Geostationary Operational Environmental Satellite} (GOES) observations for isothermal interpretation of the X-ray measurements. Temperature diagnostics of the flaring plasma have been carried out by means of a differential emission measure (DEM) analysis based on RESIK X-ray spectra.

\end{abstract}
\end{opening}
%-------------------------------------------------

\section{Introduction}
     
Long duration events (LDE) are a class of flares  characterized by slow time variations of their X-ray emission (many hours) and decay times significantly longer than the rise phase. LDEs usually occur in long-lived active regions \citep{Kahler_1977ApJ...214..891K} and their morphological structure is rather complicated. LDEs are often associated with initial large coronal mass ejections (CME) and long-duration post-eruptive energy releases  at different wavelengths \citep{Svestka_1983SSRv...35..259S}. They usually have a few hot and bright kernels of emission present along the arcade channel \citep{Jakimiec_1997AdSpR..20.2341J, Kolomanski_2002ESASP.506..665K}

The extreme ultraviolet (EUV) emission rises at different phases of the flare evolution \citep{Aschwanden_2009SoPh..256....3A}, based on the flare catalogue of the Extreme UltraViolet Imager onboard  the \textit{Solar Terrestrial Relations Observatory} (STEREO/EUVI). \cite{Howard_2008SSRV.136..67H} and \cite{Kaiser_2008SSRV.136..5k} identified at least three components of EUV emission during the flare. The first one, coincident with enhanced hard X-ray emission, is observed in 79~\%~of events during the impulsive phase. The second  component is related to the cooling of flaring loops,  which is accompanied with corresponding changes in the maximum emission from soft X-rays to EUV temperature range and was noticed during the decay phase of 73~\%~of flares, when the growth of ultraviolet emission without new flaring X-ray activity was observed. The last effect is the EUV dimming that is caused by the evacuation of expanding plasma during the initial phase of CMEs. 

\cite{Woods_2011ApJ...739...59W} based on \textit{Solar Dynamic Observatory} (SDO) observations, reported the fourth flare phase observed during the late phase of flares, which is related to development and brightening of a system of post-eruptive loops of longer length (separated in height from an original flare site).

As concers the thermal energy content of the plasma, \cite{Jiang_2006ApJ...638.1140J} reported for six short flares the thermal energy not exceeding 10$^{30}$~ergs. \cite{Bak_2011SoPh..271...75B} analysed physical conditions of flares with rise phase lasting longer than 30 minutes (so called ``slow-long" duration events, SLDEs). The total thermal energy content for the analysed flares has been determined there to be $\approx$~30$^{31}$~ergs at the maximum.                                                                                                                                                                                                                                                                                                                                                                                                                             

In this article we present a detailed analysis of a long-duration event on 15 April 2002 when the RESIK spectra were available. RESIK was an uncollimated Polish bent crystal spectrometer \citep{Sylwester_2005SoPh..226...45S} operating onboard the Russian {\em Coronas-F}  spacecraft  \citep{Kuznetsov_2014ASSL..400....1K} from 2001 to 2003. The basic components of RESIK were two double spectrometers. Each spectrometer was equipped with two silicon and quartz crystals. The bent crystal causes the radiation beam incident upon the crystal to be reflected at slightly different angles according to Bragg's law, so it was possible to obtain spectra in a selected range of wavelengths instantaneously. The RESIK spectrometer observed simultaneously spectra in four channels, which covered the spectral range from  3.3~\AA~to 6.05~\AA.  
The spectra observed in each channel were recorded at 256 spectral bins with data gathering interval (DGI) dynamically determined by the onboard computer according to the level of solar X-ray emission. 
 
RESIK observations  give a possibility of X-ray plasma diagnostics in the temperature range 3 -- 30 MK by determining the  differential emission measure distributions \citep{Kepa_2006SoSyR..40..294K, Chifor_2007A&A...462..323C, Sylwester_2008_JApA...29..147S}.

In the determination of differential emission measure distributions we used two algorithms: the Withbroe--Sylwester (WS) approach \citep{Sylwester_1980A&AS...40..335S} and the differential evolution (DE) method \citep{Storn_1997}. The DEM distributions obtained by the DE method were presented earlier by \cite{Kepa_2016IAUS..320...86K}. In the present article, we show the results of DE and WS methods as tested on selected synthetic models.

As the flare on 15 April 2002 was also observed by the \textit{Extreme Ultraviolet Imaging Telescope} (EIT) onboard the \textit{Solar and Heliospheric Observatory} (SOHO) spacecraft \citep{Delaboudiniere_2008SSRV.136..67H} and the \textit{Reuven Ramaty High Energy Solar Spectroscopic Imager} (RHESSI: \citealp{Lin_2002SoPh..210....3L}) an investigation of flare morphology and the evolution of hard X-ray emissions was possible.

This article is organized as follows: a description of instruments and observations of 15 April 2002 flare is presented in Section 2. Results of isothermal and multithermal analysis with application of WS and DE method are shown in Section 3, and the discussion is provided in Section 4.

\section{The 15 April 2002 flare}
 We studied the M1.2 GOES class flare ({\textsf{SOL2002-04-15T03:55}}) that occurred in NOAA  active region~9906 located at the central part of the solar disk (S15W01). Based on the Solar Geophysical Data report, this event started at ~03:05~UT, reached maximum at~03:55~UT, and ended at~05:06~UT. GOES light curves show that emission corresponding to this flare was observed for much longer (see Figure~1). The analysed flare was preceded by another event at 02:51 UT, looking like a precursor. However, this earlier flare was related to another active region located at the limb (see Figure~2).
\begin{figure}[h!]  %%%%%%%%%%%%%%%%%% FIGURE 2
\advance\leftskip -1.5cm{\includegraphics[width=1.5\textwidth]  {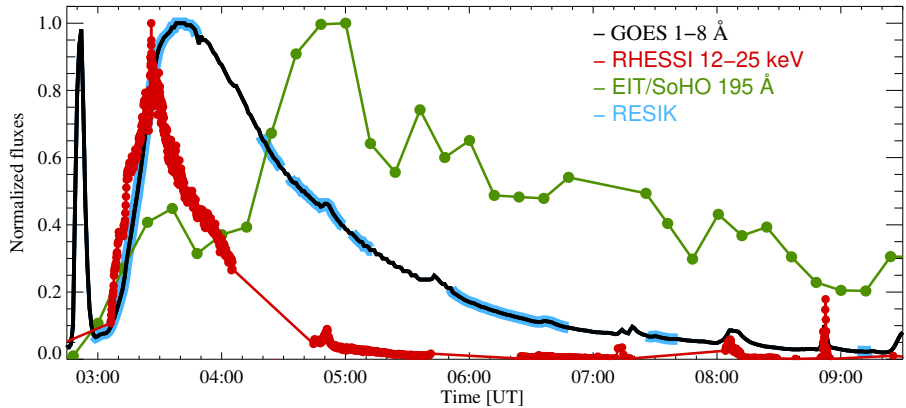}} 
\vspace{-15.3cm}
\caption{ GOES (in \textit{black}), RHESSI 12 -- 25 keV (\textit{red}) and SOHO/EIT (\textit{green}) light curves  for the \textsf{SOL2002-04-15T03:55} flare. The \textit{blue strips} on the black line (GOES lightcurve) indicate  time intervals when RESIK spectra were available.}
\label{F-simple}
\end{figure}                                                                                                                                                                                                                                                                                          

\label{S-aug}
\begin{figure}  %%%%%%%%%%%%%%%%%% FIGURE 1 
\centerline{\includegraphics[width=1.2\textwidth]{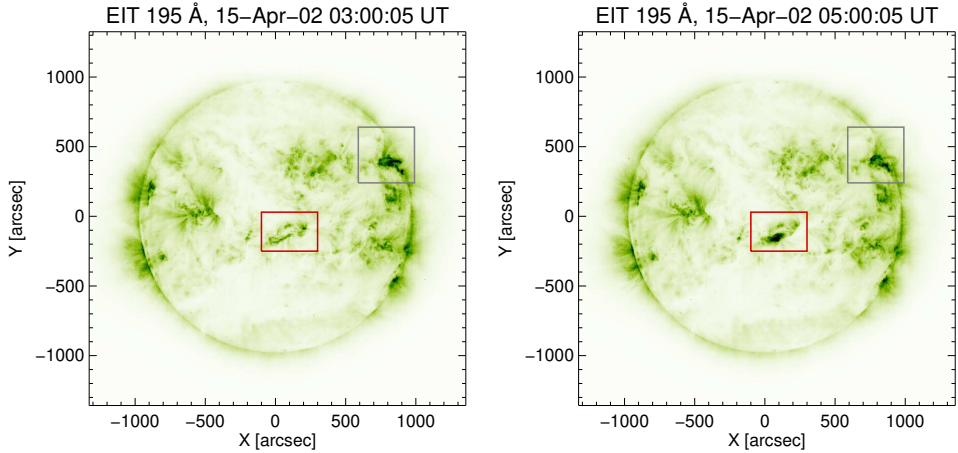}} 
\vspace{-12cm}
\caption{SOHO/EIT Sun images taken in 195~\AA~filter at two moments: 03:00:05~UT (\textit{left}) and 05:00:05~UT (\textit{right}). The \textit{red rectangle} indicates the active region in which the analysed flare occurred. The \textit{gray rectangle} is related to the location of the preceding event.}
\label{F-simple}
\end{figure}

\begin{figure}[h!]  %%%%%%%%%%%%%%%%%% FIGURE 3 
\advance\leftskip -1.5cm{\includegraphics[width=1.1\textwidth]{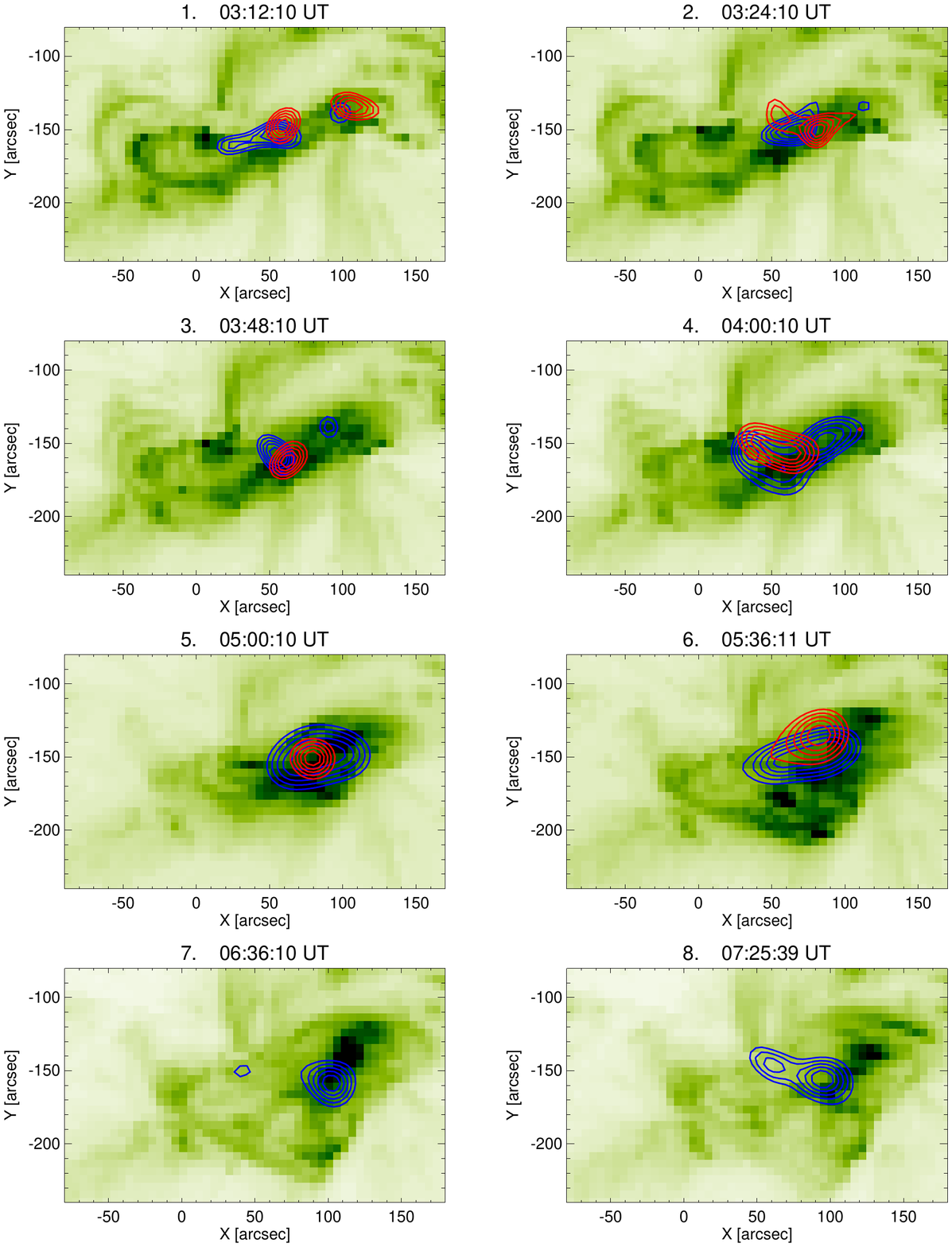}} 
\caption{The SOHO/EIT 195~\AA~images with RHESSI contours obtained in two energy ranges: 6 -- 7 keV (in \textit{blue}) and 12 -- 16 keV (in \textit{red}) are presented for selected times. The~RHESSI image reconstruction was made using the PIXON method implemented in SolarSoft (grids 3, 4, 5, 6, 8, and 9 have been used).}
\label{F-simple}
\end{figure}

Figure 1 displays the temporal profiles of the soft X-rays from GOES (1~--~8~\AA), RHESSI in the energy range 12 -- 25~keV, and the ultraviolet radiation within the area selected (red rectangle in Figure 2) in the 195~\AA~from SOHO/EIT filter. The blue  strips on the GOES light curve indicate times when RESIK observations were available. The SOHO/EIT images in 195~\AA~were taken every 12 minutes only. Respective images have been calibrated using the {\sf{eit\_prep.pro}} procedure (including corrections to the dark current, exposure time normalization, response correction,  \textit{etc.}). The EIT light curve was derived based on the integration of the signal measured in the area corresponding to the red rectangle marked in Figure~2 after background subtraction (assumed as the level of quiet Sun emission in the immediate vicinity of the considered area/region).

It is seen that the ultraviolet emission appears to have two maxima. The first slight increase of the EUV emission takes place during the rise phase of the soft X-rays and is related to peaking of hard X-rays. The second maximum, much stronger than the first one, was observed during the decay phase of X-ray flare. In this phase the SOHO/EIT light curve is not monotonic showing a few maxima (the observations were available only every 12 minutes).

A comparison of the light curves in different energy bands allows us determine the time delay between the maxima of the UV (max$_{\mathrm{uv}}$), soft (max$_{\mathrm{sxt}}$) and hard X-ray (max$_{\mathrm{hxt}}$) emissions. For the {\textsf{SOL2002-04-15T03:55}} flare the delays are as: max$_{\mathrm{hxt}}$ - max$_{\mathrm{sxt}}$ $\approx$ 20 minutes, max$_{\mathrm{hxt}}$ - max$_{\mathrm{uv}}$ $\approx$ 1 hour 30 minutes, and max$_{\mathrm{sxt}}$ - max$_{\mathrm{uv}}$ $\approx$ 1 hour 10 minutes. 

RHESSI observed almost the entire main flare phase and the three time intervals during the decay phase (see Figure 1). For {\textsf{SOL2002-04-15T03:55}} flare the emission up to 25 keV was observed. 

The RHESSI image reconstruction was made using the PIXON method \citep{Hurford_2002SoPh..210...61H} available in SolarSoft applying grids 3, 4, 5, 6, 8, and 9. The time sequence of the EIT~195~\AA~images with RHESSI contours overplotted obtained in the energy range 6 -- 8 keV and 12 -- 16 keV are presented in Figure 3. It can be observed that the configuration of flaring sources was rather complicated. The emission was still visible a few hours after maximum of the event. At the beginning of the rise phase (image No.~1 in Figure 3) two separated sources are observed,  12 minutes later (image No.~2) only one source is visible. After 04:00 UT, the source becomes much larger (images No.~4, 5, 6). At 05:00~UT (image No.~5) when the second maximum of ultraviolet emission is observed  a substantial brightening of the low lying loops system on the EIT image can be noticed. The contours of the RHESSI source overlap with the location of the source of ultraviolet emission. The last three images depict the final evolution of this emission system showing a weakening of the observed X-ray source. Starting from 06:30~UT  (three hours from X-ray maximum) only the emission at the lower energy range  (6 -- 8 keV) can be observed.
 
RESIK observations are available from 02:45 UT to 09:14 UT with a few breaks when the \textit{Coronas-F} spacecraft crossed through a polar van Allen radiation belt or it was  the South Atlantic Anomaly when the high-voltage of spectrometer was turned off. However, the first 15 minutes of observation (extending to about 03:00~UT) were associated with the earlier event (see Figure~1). 

For the {\textsf{SOL2002-04-15T03:55}} flare RESIK obtained almost 900 spectra during about 2.5 hours with cadence of ten seconds. The spectra during entire rise phase (about 35 minutes) and during more than two hours of decay phase were available. Unfortunately, due to onboard computer set-up problems the channel 3 data (for this flare) were not available. The average RESIK spectrum for {\textsf{SOL2002-04-15T03:55}} flare with strong emission lines indicated is presented in Figure~4.

\begin{figure}    %%%%%%%%%%%%%%%%%% FIGURE 4 
\centerline{\includegraphics[width=0.93\textwidth]{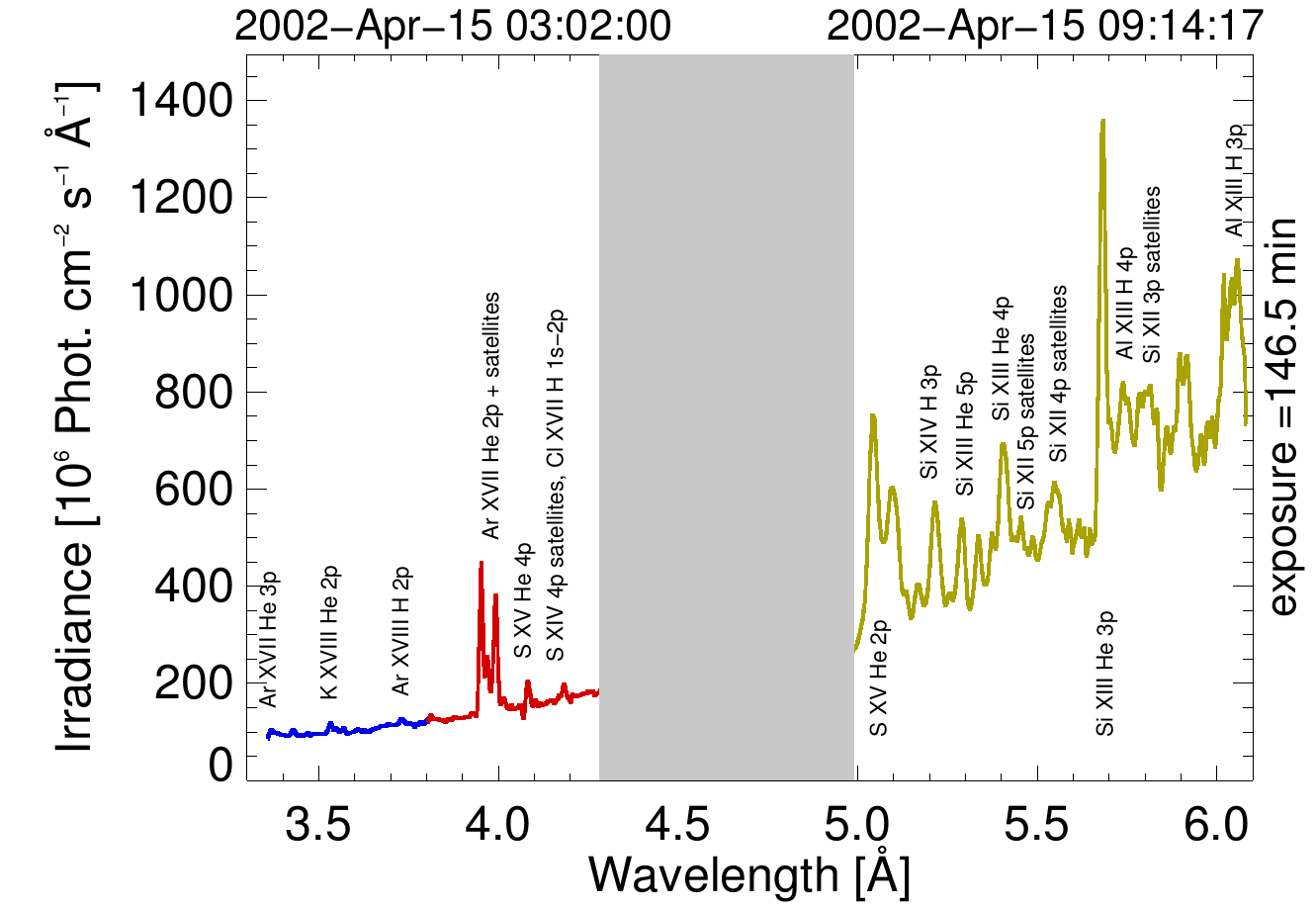}}
\vspace{.7cm}
\caption{The integrated spectrum of {\textsf{SOL2002-04-15T03:55}} flare (146 minutes of integration time) for the three RESIK channels (1: 3.3 \AA~--~3.8 \AA, \textit{blue}; 2: 3.83 \AA~--~4.3~\AA, \textit{red}; 4:~5.00~\AA~--~6.05~\AA, \textit{olive}). Due to technical problems at the time of these observations the channel 3 (4.35~\AA~-- 4.86~\AA) data are not available for the analysis.}
\label{F-simple}
\end{figure}

%%%%%%%%%%%%%%%%%%%%%%%%%%%%%%%%%%%%%%%%%%%%%%%%%%%%%%%%%%%%%%%%%%%%%%%%%%%%%%%%%%%%%%%%%%%%%%%%%%%%%%%%

\section{The Analysis}
\subsection{The Isothermal Approximation} %%%%%%%%%%%%%%
 We determined the temporal profiles of temperatures [$T_{\mathrm{e}}$] and emission measures [$EM$] based on GOES X-ray monitor data. These parameters  were derived using the isothermal approximation and applying  the flux-ratio technique. The pre-flare pedestal was subtracted from the GOES fluxes (0.5 -- 4~\AA~and 1 -- 8~\AA~channels). The calculations were made using the standard routine \textsf{\em goes.pro} available in the SolarSoft package. The flaring plasma has reached its maximum temperature, 14.6~MK, at 3:20 UT. About 50 minutes later (at 04:08 UT), the maximum of emission measure equal to 7.6$\times$10$^{48}$~cm$^{-3}$ was observed. The temporal evolution of temperature and emission measure as calculated based on GOES are presented in Figure 5 (upper-left panel).  

\begin{figure}[ht!]    %%%%%%%%%%%%%%%%%% FIGURE 1
 \vspace{-1cm}
 \centerline{\includegraphics[width=1.\textwidth]{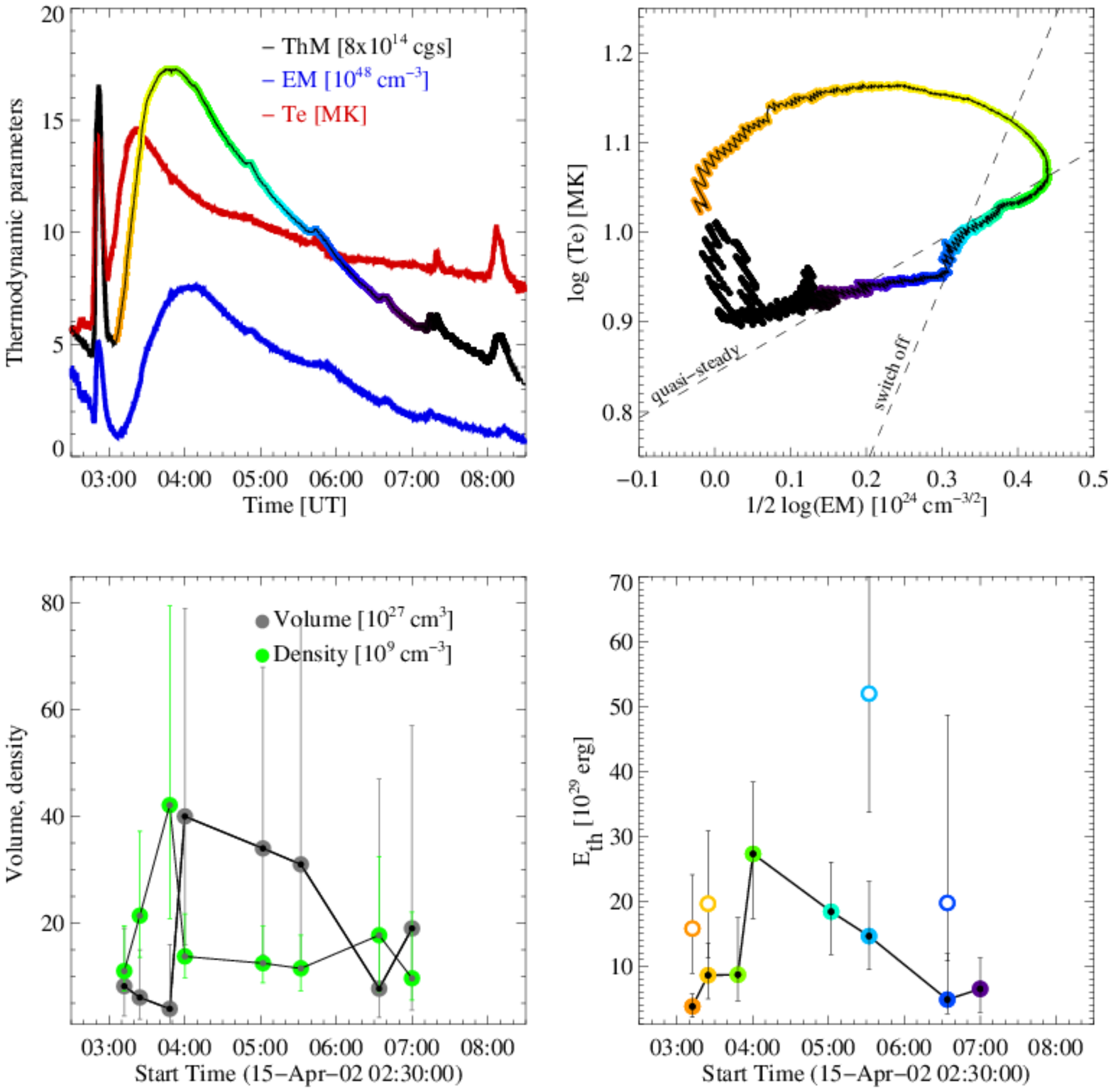}}
 \vspace{-3cm}    
 \caption{\textit{Top-left}: The temporal variations of temperature, emission measure, and the thermodynamic measure  calculated based on GOES fluxes using the isothermal approximation for the {\textsf{SOL2002-04-15T03:55}} flare, \textit{right}: the diagnostic diagram (DD) for the estimated $T_{\mathrm{e}}$ and $EM$ values. The \textit{dashed lines} on the DD plot indicate two ''limiting cases'': quasi-steady state and instant switching off of the heating. \textit{Bottom}: temporal evolution of flaring volume, average ''low density'' limit (\textit{left}) and the total thermal energy content (\textit{right}) calculated based on the volume estimated from RHESSI images reconstructed in the 6 -- 8 keV energy range. The \textit{colors in the diagnostic diagram} correspond to the colors representing the thermodynamic measure (\textit{upper-left panel}). \textit{Circles with a black dot} represent thermal energy obtained for isothermal approximation, \textit{empty circles}  represent $E_{\mathrm{th}}$ values calculated based on the differential emission measure distributions and constant pressure assumption (see Equation 6 and the discussion in Sectopm 3.2.3)}.
   \label{F-simple}
   \end{figure}

An important parameter characterizing the total accumulated energy of flare plasma is the thermal energy ($E_{\mathrm{th}}$). By definition for the isothermal plasma of density $N_{\mathrm{e}}$ and volume $V$ it is:
\begin{equation}
 E_{\mathrm{th}}=3{\mathrm{k}}T_{\mathrm{e}}N_{\mathrm{e}}V=3{\mathrm{k}}T_{\mathrm{e}}\sqrt{EM}\sqrt{V}~~~~~~~~~~~~~~~  \mathrm{[erg]}
\end{equation}
where k is Boltzmann constant and $EM=N_\mathrm{e}^2V$.

Provided the flaring plasma volume is constant, we can use so-called thermodynamic measure ({\em ThM}), \cite{Sylwester_1995A&A...293..577S}. 

\begin{equation}
ThM=3{\mathrm{k}}T_{\mathrm{e}}\sqrt{EM}~~~~~~~~~~~~~~~~~~~{\mathrm{[g~cm^{1/2} s^{-2} ]}}
\end{equation}

The {\em ThM} represents the thermal-energy content in the X-ray plasma source. 
\begin{equation}
E_{\mathrm{th}}=\sqrt{V}~ThM~~~~~~~~~~~~~~~~~~~~~  {\mathrm{[erg]}}
\end{equation}

Thermal energy can also be determined for multi-temperature plasma assumption. Appropriate equations are presented in the next section.
Respective temporal variations of thermodynamic measure for the analysed flare are presented on the upper left panel in  Figure 5. 

For the eight time intervals, RHESSI images have been reconstructed (see Figure~3), so it was possible to estimate the plasma volume by assuming spherical symmetry of emitting source, calculate the lower limit of plasma density (filling factor assumed = 1) and determine the total thermal energy content. The plasma volume was estimated using  RHESSI images obtained in the 6 -- 8~keV energy range (the 50~$\%$ contour was used in this respect). For the assumed shape of the source, the estimated volumes changed from $3.9\times10^{27}$ ~cm$^3$ to $4\times10^{28}$~cm$^3$. The respective values for the electron density of the hot plasma are $9.6\times10^{9}$~ cm$^{-3}$ and $4.2\times10^{10}$~cm$^{-3}$.

The amount of thermal energy with the above assumptions is of the order of 10$^{30}$ ergs (see the lower panel of Figure~5). In order to estimate errors in the calculated  density and thermal energy, the volumes of the source were determined additionally for contours 30~$\%$ and 70~$\%$. For the volume of source defined in this way, the obtained values of density and thermal energy differ from those obtained for contour 50~$\%$  by a factor of 2 and  1.5, respectively.

Based on the results of hydrodynamic modelling  of the flare (code Palermo-Harvard: PH),  \cite{Jakimiec_1992A&A...253..269J} proposed the method for a heating and cooling diagnostic of a flaring loop. This method is based on the analysis of the trajectory slopes on the density -- temperature diagnostic diagram  (DD). Two limiting slopes of flare decay phase were identified that correspond to the situation of a sudden switch off of the flare heating and the evolution along a so-called quasi-steady state line. Intermediate inclinations correspond to the decrease of flaring heating with a decay time $\tau$ [$E_{\mathrm{H}}(t)\sim {\mathrm{e}}^{\frac{-t}{\tau}}$, where $E_{\mathrm{H}}$ is the flare heating rate]. 
In this article we used a simplified version of diagnostic diagram in which the density is replaced by the square root of the emission measure (\textit{i.e.} constant volume of the emitting source is assumed). 

Based on the inspection of the diagnostic diagram  the timing of the nearly constant heating [$\Delta t$]  has been determined. For the {\textsf{SOL2002-04-15T03:55}} flare $\Delta t$ was about one hour. Subsequently the evolution of the flare was along the quasi-steady state branch. 
 
One must remember that the PH hydrodynamic modelling  was performed with the following, rather strong, assumptions:
\begin{itemize}
\item 
the flaring plasma is confined in a single loop of constant cross section. 
\item
the uniform (across the structure) heating is located at the loop top.
\end{itemize}

In the case of the investigated flare, the morphology is rather complicated, so the results obtained using the diagnostic diagram must be regarded as ''indicative''. 
 
\subsection{Determination of the Differential Emission Measure Distribution} %%%%%%%%%%%%%% 

\subsubsection{Methods}
The differential emission measure [$\varphi(T_{\mathrm{e}})$] may loosely be regarded as the amount of the emitting plasma present in a given temperature range (for a physical interpretation of the DEM, see, \textit{e.g.} \citealp{Sylwester_1980A&AS...40..335S, Guennou_2012ApJS..203...26G}):
\begin{equation}
\varphi(T_{\mathrm{e}})=N_{\mathrm{e}}^2\frac{{\mathrm{d}}V}{{\mathrm{d}}T_{\mathrm{e}}}              
\end{equation}
where: $N_{\mathrm{e}}$ is electron density, $V$ is plasma volume, and $T_{\mathrm{e}}$ is temperature.
Its form can be determined by solving a set of integral equations defining the observed line fluxes.
 \begin{equation}
 F_i=A_i\int_0^\infty  f_i(T_{\mathrm{e}})\varphi(T_{\mathrm{e}})\, {\mathrm{d}}T_{\mathrm{e}}~~ i=1,2, ... N
 \end{equation}
where  $A_i$ represents the assumed abundance of an element contributing to the flux of a particular line or spectral interval, $N$ is the number of spectral bands used, and $f_i(T_{\mathrm{e}})$ is the theoretically calculated emission function for a given line/band.
The emission function represents the flux in the selected line/band at the temperature $T_{\mathrm{e}}$  for a given set of abundances at a distance of 1 AU. The emission functions depend on the adopted calculations of the ionization equilibrium  and the populations of excited levels associated with the formation of the line.

The idea of determining the DEM is to find such a profile of DEM distribution that provides the most probable agreement between observed and  calculated based on  this distribution fluxes (the right side of Equation 5). 

Both the emission functions and the observed fluxes have uncertainties. It is difficult to estimate the uncertainties of the emission  functions. Therefore, it is usually assumed that their values are determined accurately and uncertainties in the resulting calculated distributions of DEM are due to uncertainties in the observed fluxes only.

The reconstruction of differential emission measure distributions  is a well-known example of an ill-posed mathematical problem. A direct inversion of the data does not produce a unique DEM solution and additional constraints are needed to achieve a stable solution.

Nowadays, there are a number of methods for inversion of  DEM shape from solar and stellar data -- see for example \cite{Aschwanden_2015SoPh..290.2733A}. In the present study we show  DEM distributions  obtained using the two algorithms: Withbroe -- Sylwester (WS) and  differential evolution (DE).

The Withbroe--Sylwester (maximum likelihood) multiplicative algorithm is an iterative procedure that allows for determination of DEM shape without setting additional conditions on the character of the distribution, and its solution is always positive \citep{Fludra_1986SoPh..105..323F}.

The second method that we used here is based on the  mechanisms of differential  evolution (DE). Differential evolution \citep{Storn_1997} is a stochastic evolutionary algorithm used for global optimization. DE is a population based  algorithm  like  genetic  algorithms  using  similar  operators: crossover,  mutation,  and  selection. The distributions of differential emission measure obtained based on genetic algorithm for {\em Hinode} data were presented  by \cite{Siarkowski_2008AnGeo..26.2999S}.  The  main  difference in obtaining a final solution is that  genetic algorithms rely on crossover  while DE relies on the mutation operation.  In the DE method the mutation is based on the differences of randomly sampled pairs of solutions in the population \citep{Storn_1997}.
Here, we adopted this approach for the calculation of DEM distributions. DE works with a set of randomly generated individuals (corresponding to (in our case) DEM distributions), representing possible solutions of the problem.

\subsubsection{Tests of DEM Inversions}  
In our previous article \citep{Kepa_2016IAUS..320...86K} we analysed DEMs for another flare using for the first time the DE approach. In the present work, we reduced the number of temperature intervals and calculate the DEM distributions in the temperature range from 3 to 30 MK.   
 
Based on RESIK spectrum of {\textsf{SOL2002-04-15T03:55}} flare (see Figure 4) we have chosen 17 spectral bands covering the respective wavelength intervals. Their main characteristics (wavelength ranges and the most important line contributions) are presented in Table~1. The corresponding emission functions for these spectral bands containing both line and continuum contributions were calculated using the CHIANTI v. 8.1 atomic data package with  ionization equilibrium from \cite{Bryans_2009ApJ...691.1540B}.  
For silicon, sulfur, argon, and potassium we used abundances calculated using the {\sf{AbuOpt}} method from multithermal assumption \citep{Sylwester_2015ApJ...805...49S}. We assumed $A_{\mathrm{Si}}$=4.07$\times10^{-5}$, $A_{\mathrm{S}}$=2.13$\times10^{-6}$, $A_{\mathrm{Ar}}$=5.49$\times10^{-6}$, $A_{\mathrm{K}}$=2.95$\times10^{-7}$. For other elements we adopted abundances from {\sf{sun\_coronal\_ext.abund}} (available in the Chianti 8.1 package).  Figure~6 (right panel) displays the temporal evolution of RESIK fluxes observed in selected wavelength ranges and respective, theoretically calculated, emission functions (left panel). These data sets were used in determinations of differential emission measure distributions.

\begin{table}[ht!]
  \caption{Spectral bands used in calculation of DEMs.}
 \begin{tabular}{r c c c}
{\bf No.} & {\bf Wavelength range [\AA]} & {\bf Main line }  & {\bf RESIK channel } \\ 
\hline
 {\bf 1} & {3.454 -- 3.466} & {Ar {\sc{xvi}} 3p}  & {1}\\ 
 {\bf 2} & {3.520 -- 3.542} & {K {\sc{xviii}} (w)}  & {1} \\
 {\bf 3} & {3.542 -- 3.556} & {K {\sc{xviii}} (x+y)}  & {1}\\ 
 {\bf 4} & {3.745 -- 3.759} & {Ar {\sc{xviii}} 2p}  & {1}\\ 
 {\bf 5} & {3.941 -- 3.961} & {Ar {\sc{xvii}} 2p (w)}  & {2}\\ 
 {\bf 6} & {3.961 -- 3.977} & {Ar {\sc{xvii}} 2p (x)}  & {2} \\ 
 {\bf 7} & {3.977 -- 4.003} & {Ar {\sc{xvii}} 2p (z)} & {2} \\ 
 {\bf 8} & {4.093 -- 4.109} & {S {\sc{xv}} 4p (z)}  & {2} \\ 
 {\bf 9} & {5.009 -- 5.077} & {S {\sc{xv}} 2p (w)} & {4}\\ 
 {\bf 10} & {5.078 -- 5.120} & {S {\sc{xv}} 2p (x+y)}  & {4}\\ 
 {\bf 11} & {5.198 -- 5.242} & {Si {\sc{xiv}} 3p} & {4} \\ 
 {\bf 12} & {5.257 -- 5.316} & {Si {\sc{xiii}} 5p}  & {4}\\ 
 {\bf 13} & {5.376 -- 5.426} & {Si {\sc{xiii}} 4p}  & {4}\\ 
 {\bf 14} & {5.526 -- 5.584} & {Si {\sc{xii}} sat.}  & {4}\\ 
 {\bf 15} & {5.659 -- 5.709} & {Si {\sc{xiii}} 3p}  & {4}\\
 {\bf 16} & {5.782 -- 5.845} & {Si {\sc{xii}} 3p sat}  & {4}\\   
 {\bf 17} & {6.046 -- 6.063} & {Al {\sc{xiii}} 3p}  & {4}\\   
\end{tabular}
\vspace{3mm}
 \end{table}
 
\begin{figure}[h!]    %%%%%%%%%%%%%%%%%% FIGURE 1 
   \centerline{\includegraphics[width=1.1\textwidth]{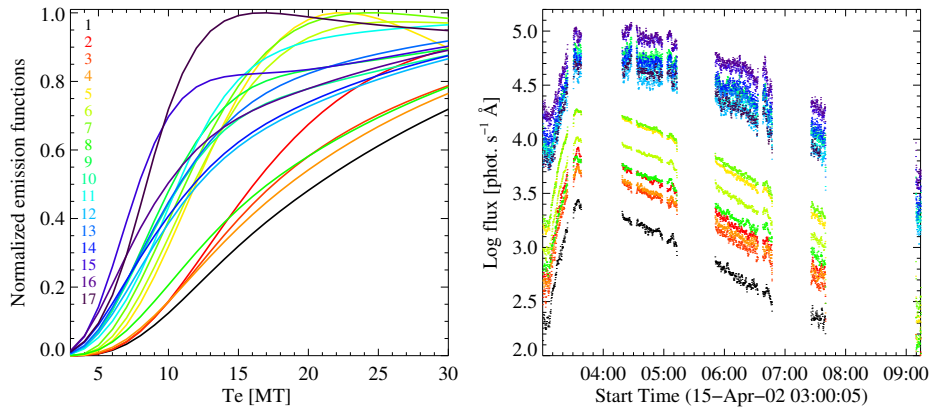}}
   \vspace{-10.5cm}
            \caption{ \textit{Left}: The emission functions for selected 17 wavelength ranges used in the present study. \textit{Coloured key numbers} denoting the curves correspond to bands given in Table 1. \textit{Right}: The temporal evolution of RESIK fluxes observed at selected wavelength ranges for the investigated flare. These fluxes were used as input data for derivation of differential emission measure distributions.}
   \label{F-simple}
\end{figure}

\begin{figure}[ht!]
 \centerline{\includegraphics[width=1.2\textwidth]{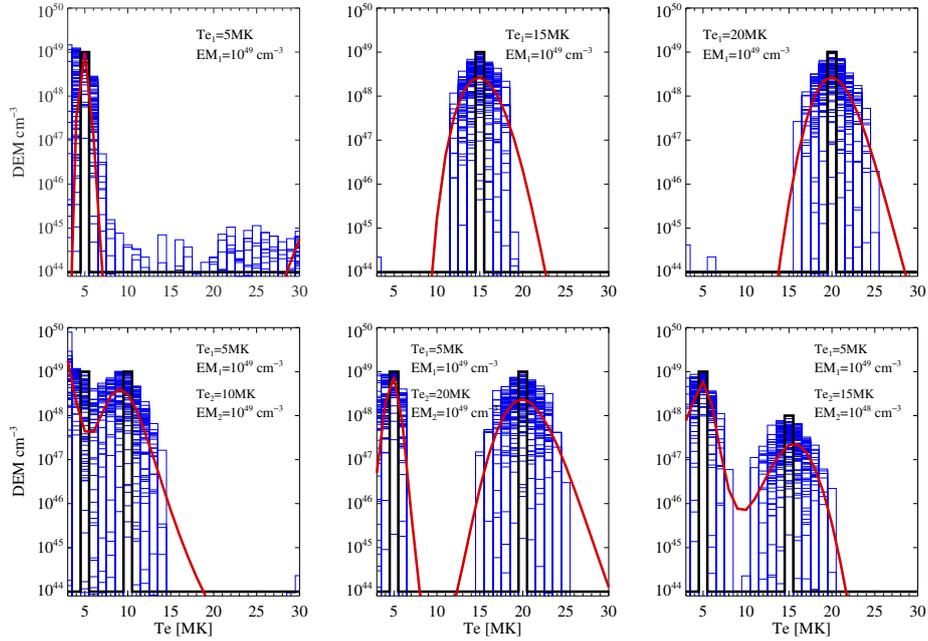}} 
 \vspace{-9.5cm}
     \caption{The comparison of assumed (\textit{black line}) with 100 calculated (\textit{blue rectangles}) models obtained using the differential evolution (DE) method and Withbroe--Sylwester (WS) approach (in \textit{red curves}).  \textit{Upper panel}: results for isothermal DEM models at 5, 15 and 20 MK; \textit{lower panel}: for two-temperature synthetic models: the same amount of plasma at 5 MK and 10 MK, 5 and 20 MK, and with different amount of plasma at 5 and 15 MK.}
\label{F-simple}
 \end{figure} 

\begin{figure}[p]    %%%%%%%%%%%%%%%%%% FIGURE 1 
 \centerline{\includegraphics[width=1.5\textwidth]{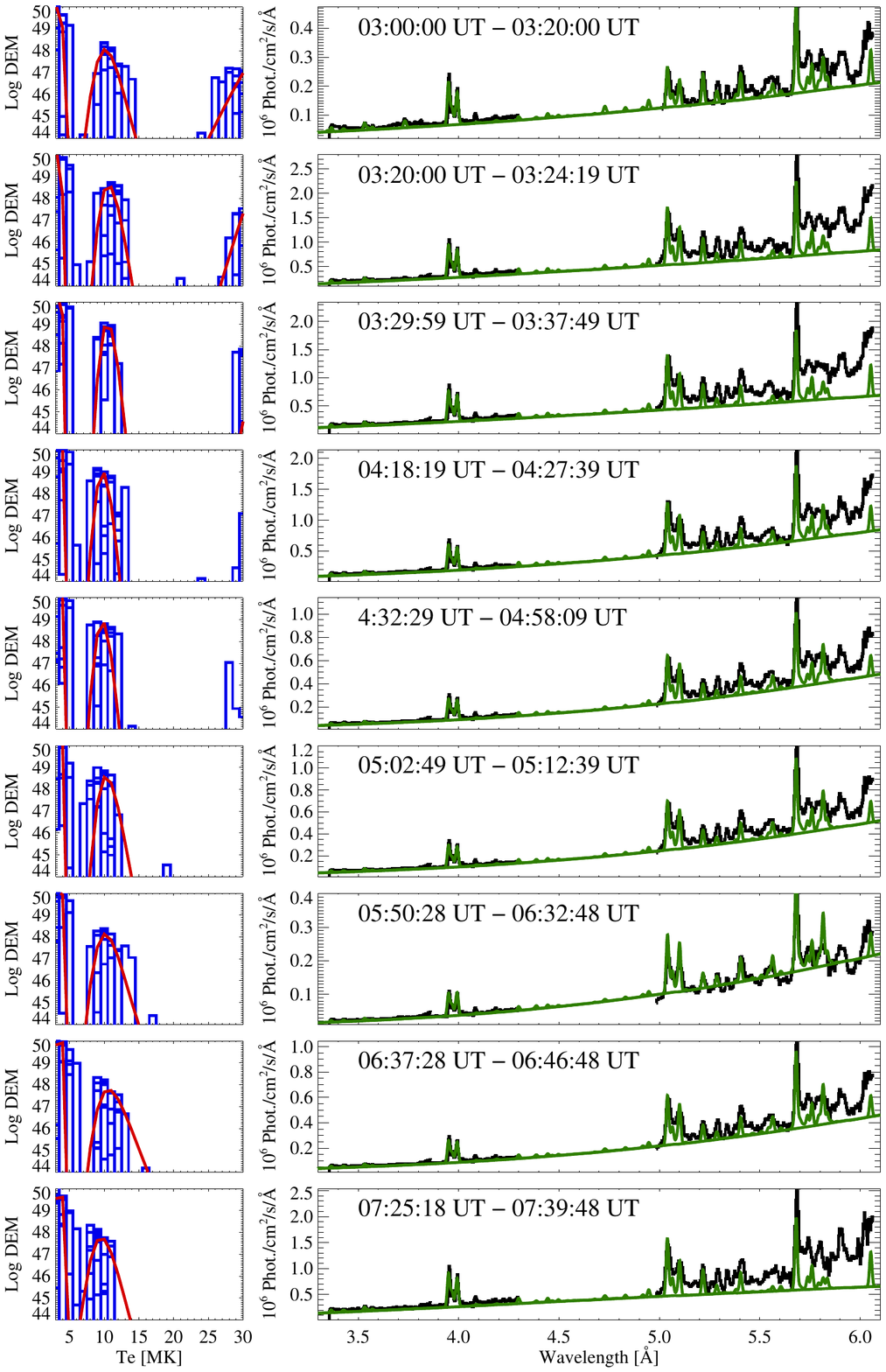}}
 \vspace{-3cm}
   \caption{ The  sequence  of derived  DEM  distributions (\textit{left panel}) and RESIK measured and calculated spectra (\textit{right}) taken at nine selected temporal intervals. 
\textit{Each  row}  corresponds  to  one  temporal  interval. \textit{Colors blue and red} are related to  difference evolution (DE) and Withbroe--Sylwester (WS) inversion methods, respectively. For each temporal interval for DE tem differential emission measure distributions correspond to ten independent generations of population  are presented. In the \textit{right panel} the RESIK observed spectra are presented in \textit{black}. Synthetic spectra (in \textit{green}) were calculated based on best fitted DEM distributions obtained using DE method shown in the \textit{left panel}.}
   \label{F-simple}
 \end{figure}

 To validate the success of reconstruction  and assess the quality of DEM calculation we performed a number of tests of the inversion methods used in the present article. In this respect we assumed the shape of synthetic model of differential emission measure and calculated (from Equation~ 7) fluxes in the set of basic bands indicated in Table 1. Then each of calculated fluxes was perturbed by random error (15~\% of the value) and we treated the calculated set as the observed one ($F_{\mathrm{obs}}$=$F_{\mathrm{calc}}$+$\varepsilon$, where $\varepsilon$ is a random error). Next we used this set of values as an input data for the inversion. The results of tests for six selected iso- and multithermal models of DEM for WS and DE approaches are presented in Figure 7. 
 
The black line represents the assumed synthetic model of DEM and blue lines illustrate
100 results as obtained using the differential evolution method. To do this, each of the flux values has been perturbed 100 times, so 100 different sets of the input fluxes and 100 results were obtained. The  process  of  evolution  for each restored model was  stopped  after 10,000 iterations,  when the convergence expressed in terms of $\chi^2$ became very slow. 

For clarity, the results  of reconstructions the synthetic models of DEM using the Withbroe -- Sylwester method (in red) are presented only for the unperturbed fluxes.

One can see that both DE and WS methods reconstruct satisfactorily the isothermal and two-temperature models.

\subsubsection{DEM Distributions for 15 April 2002 Flare}

Based on RESIK light curves we selected nine temporal intervals for detailed DEM reconstruction. The  timings are given in Figure 8 in the right column. For each one we calculated the mean spectrum and fluxes in 17 sets of wavelength ranges (see Table 1). To avoid the contribution of non-flaring plasma, the preflare X-ray fluxes have been subtracted.  The differential emission measure distributions have been determined using two method described above: Withbroe--Sylwester (WS)  and differential evolution (DE).  The calculations have been performed in the temperature range 3 -- 30 MK. The results are presented in the left panel of Figure~8. In the right panel  we show the comparison of our best fit calculated spectrum with observations. We calculated spectra based on DEM distributions obtained using the DE method and CHIANTI ver. 8.0.1. We used the same ionization balance and set of element abundances  as we used in the calculations of emission functions. 
The discrepancy between observed and synthesized spectra in spectral range 5.9 -- 6~\AA~ may be associated with the contribution of emissions from higher orders of reflection in RESIK spectra, to the first-order spectra we observe in the fourth channel. The number of counts at the single detector for each position is equal to the sum of counts from the first order (at emitted wavelength $\lambda$), second order (at $\lambda$/2), third order ($\lambda$/3) \textit{etc}.  For the Si 111 cut mono crystal reflections, used in RESIK for channel~4, the second order of reflection is prohibited, while the allowed third order  of reflection includes the He-like Fe line complex (at 1.85~\AA) and Ni line complex (at 1.55~\AA). The intensity of those lines can influence on the emission spectra in channel~4. Unfortunately those effects are very difficult to account for and the work on this problem is still in progress.

The calculated DEM distributions are two-component independent of the evolutionary phase. Both WS and DE methods give  similar results. The small amount of hotter plasma (25 -- 30 MK) is seen at the beginning (rise and maximum) of this long-duration event \citep{Tanaka_1986Ap&SS.118..101T}. 

The temporal evolution of the three components of the differential emission measure distribution is shown in Figure 9. The green, blue, and red plots represent the total emission measure calculated in temperature ranges 3 -- 6~MK, 7 -- 16~MK and 25 -- 30~MK respectively.

It can be noticed that the temporal behavior of the coldest component (green color) is very similar to the GOES lightcurve shape in the 1 -- 8~\AA~range. This component is mainly determined by the emission observed in the fourth RESIK channel, where silicon spectral lines are observed. The temporal evolution of hot and hotter components also resemble flare profiles, but they are not correlated with GOES or RHESSI light curves. The hotter component is observed from the rise phase, at maximum, and up to an hour after the maximum of X-ray emission. During the  decay phase of the flare, the total amount of plasma associated with this component is almost two orders smaller than just before the maximum.

 \begin{figure}[h!]    %%%%%%%%%%%%%%%%%% FIGURE 1 
 \centerline{\includegraphics[width=1.5\textwidth]{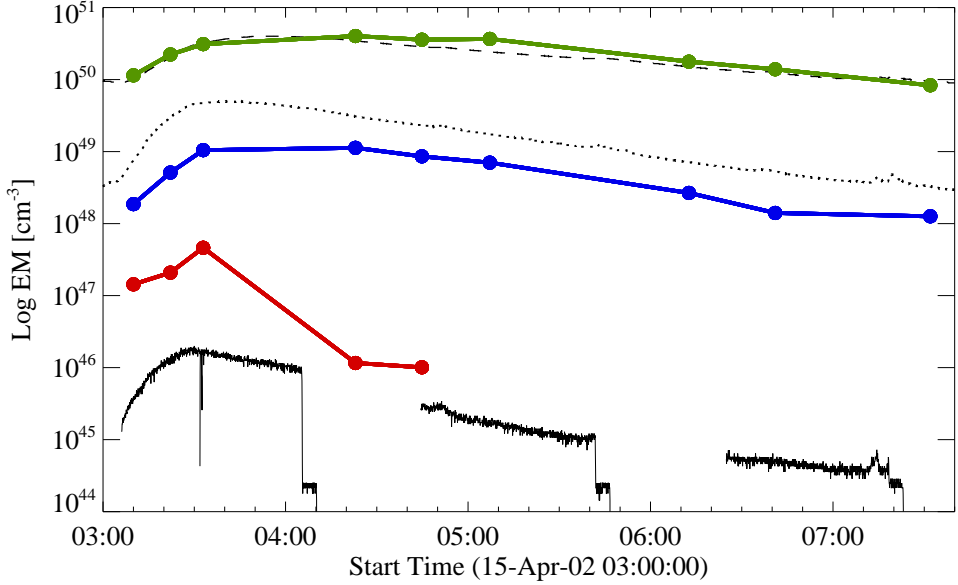}}
 \vspace{-14cm}
   \caption{The temporal evolution of the three components of the \textbf{differential emission measure} as calculated based on average DEM distributions using DE and WS methods. The \textit{green}, \textit{blue} and \textit{red} represent total emission measure present in the temperature ranges 3 -- 6~MK, 7 -- 16 MK, and 25 -- 30 MK respectively. The \textit{thin dashed} and \textit{dotted lines} present the GOES light curves in the 1 -- 8~\AA~and 0.5 -- 4~\AA~ranges. The \textit{bottom black line} is the RHESSI lightcurve in the 6 -- 12 keV energy range. The GOES and RHESSI light curves have been scaled for better visualisation. }
   \label{F-simple}
 \end{figure}   

Based on the calculated  differential emission measure distributions and the assumptions of  constant pressure (see Equation 6) or constant plasma density (see Equation 7) the thermal energy content can be calculated. 

\begin{equation}
E_{{\mathrm{th}|p=const}}=3{\mathrm{k}}\sqrt{V}\sqrt{\int{T^2\varphi(T_{\mathrm{e}}){\mathrm{d}}T_{\mathrm{e}}}}
\end{equation}

\textbf{\begin{equation}
E_{{\mathrm{th}}|N_{\mathrm{e}}={\mathrm{const}}}=3{\mathrm{k}}\sqrt{V}\frac{\int{T_{\mathrm{e}}\varphi(T_{\mathrm{e}}){\mathrm{d}}T_{\mathrm{e}}}}{\sqrt{\int \varphi(T_{\mathrm{e}}){\mathrm{d}}T_{\mathrm{e}}}}
\end{equation}}

We determined the thermal energy content for four temporal intervals for which  RHESSI images and RESIK observations were available. Our calculations have been performed for both constant pressure and constant density assumptions.  The results obtained are similar, so $E_{\mathrm{th}}$ values calculated for the constant pressure assumption have been used only. Obtained results are presented in Figure 5 (bottom-right panel). The values are higher than for an isothermal model, but the results agree within an order of magnitude ($10^{30}$ ergs).

\section{Concluding Remarks} %%%%%%%%%%%%%%%%%%%%%%%%%%%%%%%%%%%%%%%%
 Based on the data from  GOES, RHESSI, SOHO/EIT, and RESIK we carried out analysis of the long duration event that occurred on 15 April 2002 ({\textsf{SOL2002-04-15T03:55}}). We determined physical characteristics of flaring plasma (temperature, emission measure, thermodynamic measure, density, thermal energy and differential emission measure distributions) and investigated their temporal evolutions based on GOES, RHESSI, SOHO/EIT, and RESIK data. Our analysis can be summarized as follows:
\begin{itemize}
\item 
The ranges of flaring plasma temperature and emission measure calculated in an isothermal approximation based on GOES fluxes are 7.9~MK~-~14.6~MK and 1$\times$10$^{48}$~--~7.6$\times$10$^{48}$~cm$^{-3}$ respectively.
\item  The size of the source is variable during the flare evolution. From  RHESSI images reconstructed in the energy range  6 -- 8 keV and assuming the equivalent spherical shape of the X-ray source, the calculated volume changes from:  3.9$\times$10$^{27}$ cm$^3$ to 4.0$\times$10$^{29}$ cm$^3$.
\item The corresponding range of densities for the hot plasma component is 9.6$\times$10$^{9}$~--~ 4.3$\times$10$^{10}$ cm$^{-3}$.
\item The differential emission measure distributions were calculated based on two methods: Withbroe--Sylwester and differential evolution. Both of the methods were tested. The results of the tests confirmed the stability of the solutions and capability to reconstruct the synthetic distributions  in the specified temperature range from 3 to 30 MK. Both WS and DE methods provide similar DEM inversions. 
\item The obtained DEM distributions are always two-component independent of the evolutionary phase. A the small amount (the third component) of hotter plasma (25 MK -- 30 MK) is seen at the beginning  (rise and maximum) of this long-duration event. This component may be associated with the presence of suprathermal electrons  based on the coincidence with the harder X-rays.
\item The temporal evolution of the coldest component (green color in Figure 9) mimics the evolution of the GOES flux in the 1 -- 8~\AA~range.
\item The amount of thermal plasma energy content is of the order of 10$^{30}$ ergs. The values calculated by the assuming isothermal plasma model are lower than those calculated based on the differential emission measure distributions.

\end{itemize}
\begin{acks}
 We acknowledge financial support from the Polish National Science Centre grants No. 2017/25/B/ST9/01821 and 2015/19/ST9/02826.)
\end{acks}

\section*{Disclosure of Potential Conflicts of Interest}
The authors declare that they have no conflicts of interest.
\bibliographystyle{spr-mp-sola}
\bibliography{sola_bibliography_ak}
\addcontentsline{toc}{section}{\numberline{}References}
\end{article} 
\end{document}